\documentclass{ws-procs9x6}

\usepackage{epsfig,amssymb}

\usepackage{prprntno}

\newcommand{\be}{\begin{eqnarray}}
\newcommand{\ee}{\end{eqnarray}}
\newcommand{\nn}{\nonumber\\}
\newcommand\sfrac[2]{{\textstyle \frac{#1}{#2}}}  

\newcommand{\La}{{\mathcal L}}

\newcommand{\re}{{\mathrm{Re}\,}}
\newcommand{\im}{{\mathrm{Im}\,}}

\newcommand{\lessthansquiggle}{%
 \raise.3ex\hbox{$<$\kern-.75em\lower1ex\hbox{$\sim$}}}
\newcommand{\greaterthansquiggle}{%
 \raise.3ex\hbox{$>$\kern-.75em\lower1ex\hbox{$\sim$}}}

\begin{document}

\title{Standard Model and Beyond}

\author{A. Bartl and S. Hesselbach} 

\address{Institut f\"ur Theoretische Physik, Universit\"at Wien,
  A-1090 Vienna, Austria}

\begin{flushright}
\ttfamily
  UWThPh-2004-43 \\
  hep-ph/0412419
\end{flushright}
\vspace{-10mm}

\maketitle

\abstracts{
We first discuss the basic features of electroweak 1-loop corrections 
in the Standard Model.
We also give a short and elementary review on Higgs boson searches, 
grand unification, supersymmetry and extra dimensions.}

\section{Introduction}

The Standard Model (SM) is our present theory of the fundamental
interactions of the elementary particles. It includes quantum
chromodynamics (QCD) as the theory of the strong interactions and the
Glashow-Salam-Weinberg (GSW) theory as the unified theory of the
electromagnetic and weak interactions. Both QCD and GSW theory are
non-Abelian gauge theories, based on the principle of local gauge
invariance. The gauge symmetry group of QCD is SU(3) with colour as
the corresponding quantum number, that of the GSW theory is
$\mathrm{SU}(2) \times \mathrm{U}(1)$ with the quantum numbers weak
isospin and hypercharge. The gauge symmetry group of the GSW theory is
spontaneously broken by the Higgs mechanism from 
$\mathrm{SU}(2) \times \mathrm{U}(1)$ to the electromagnetic
U(1).\cite{nachtmann}
According to the gauge symmetry groups there are eight massless
gluons mediating the strong interactions, one massless photon for the
electromagnetic interaction and three vector bosons $W^\pm$ and $Z^0$
for the charged and neutral weak interactions. The weak vector bosons
aquire their masses by the spontaneous breaking of the electroweak
symmetry group.

The matter particles have spin $\frac{1}{2}$ and are grouped into
three families of quarks and leptons. The fermions appear as
left-handed and right-handed states, except for the neutrinos which in
the SM are only left-handed and massless. The left-handed fermions are
grouped in isodoublets, the right-handed fermions are isosinglets. The
quark generations are mixed by the charged weak currents. This quark
mixing is described by the Cabibbo-Kobayashi-Maskawa (CKM)
matrix. The Glashow-Iliopoulos-Maiani (GIM) mechanism
guarantees the absence of flavour changing neutral currents (FCNC) at
tree level. In QCD the coupling of the gluon to
the quarks is flavour independent (``flavour-blind''). The flavour
dependence in the SM is essentially due to the quark mixing. To
emphasize this aspect of the flavour dependence this part of the GSW
theory is also called quantum flavour dynamics (QFD). Note that in the
present formulation of the SM there is no mixing between lepton
families. While this is true to a high accuracy for the charged
leptons, we know that it is not true for the neutrino sector because
neutrino oscillations occur.\cite{petcovTalk}

The spontaneous breaking of the electroweak symmetry is achieved by
introducing one doublet of complex scalar Higgs fields. This is the
minimum number of Higgs fields necessary to spontaneously break the
$\mathrm{SU}(2) \times \mathrm{U}(1)$ symmetry and to introduce the
mass terms for all particles apart from the neutrinos. After
spontaneous symmetry breaking there remains one neutral scalar Higgs
particle as physical state. The other three scalar fields become the
longitudinal components of the massive $W^\pm$ and $Z^0$ bosons.

The SM is phenomenologically very successful. Highlights of the
experimental development were the discoveries of the $W^\pm$ and $Z^0$
bosons, the $\tau$ lepton, the heavy quarks and the gluon at the large
accelerator centres CERN and DESY in Europe, BNL, FNAL and SLAC in the
USA. At present the SM can reproduce all accelerator-based
experimental data. The gauge sector of the SM has been extremely well
tested. If radiative corrections are included, the theoretical
predictions are in very good agreement with the data of LEP, SLC,
Tevatron and HERA.\cite{giacommelliTalk} Some observables have been
measured with an error of less than one per mille, the theoretical
predictions have a similar accuracy. However, the Higgs sector has up
to now not been sufficiently well tested. In particular, the Higgs
boson has not been found yet. Our theoretical ideas about the
spontaneous electroweak symmetry breaking have still to be
verified. If the Higgs mechanism of the SM is the right way of
electroweak symmetry breaking, then we know from the direct searches
at LEP that the mass of the Higgs boson has the lower bound
$m_h > 114.4$~GeV.\cite{giacommelliTalk,higgsbound}

Despite its phenomenological success it is generally believed that the
SM is just the low-energy limit of a more fundamental
theory. Obviously, the SM in its present form cannot describe the
recent experimental results on neutrino oscillations, which are only
possible if the neutrinos have mass. Several theoretical ideas have
been proposed for introducing neutrino mass terms. For a review we
refer to Ref.~\refcite{petcovTalk}.

We have also theoretical arguments for our believe that the SM has to
be extended. One attempt is to embed the SM into a grand unified
theory (GUT) where all gauge interactions become unified at a
high scale $M_\mathrm{GUT} \approx 10^{16}$~GeV. Another extension of
the SM is provided by supersymmetry (SUSY), which is probably the most
intensively studied one so far. Other modifications are composite
models, technicolour, strong electroweak symmetry breaking, little
Higgs etc. In recent years the idea of ``large extra dimensions'' has
been proposed and intensively studied, which could also provide a
solution of some of the theoretical flaws of the SM. All these
extensions of the SM will be probed at the Large Hadron Collider
LHC,\cite{LHC} which is presently under construction at CERN and will
start operating in the year 2007. In the last decade the design of an
$e^+ e^-$ linear collider has been intensively studied.\cite{lincoll}
At such a machine all extensions of the SM could even be more
precisely tested.

In this series of lectures we will first review the basics of
electroweak radiative corrections in the SM and then present a short
comparison with the experimental data. Then we will briefly discuss
how to search for the Higgs boson in $e^+ e^-$ collisions and at
$p\bar{p}$ and $pp$ colliders. In the following sections we will
discuss some aspects of physics beyond the SM. We will shortly treat
GUTs, then give a phenomenological introduction to SUSY and close with
some remarks about large extra dimensions.

\section{Standard Model Physics}

The SM is a renormalizable quantum field theory because
QCD and the GSW theory are gauge theories.
This enables us to calculate the
theoretical predictions for the various observables with high
accuracy. In the last years both the QCD and the electroweak 1-loop
corrections for all important observables have been calculated. For
some observables even the leading terms of the higher order
corrections are known. Moreover, also the QCD corrections to a number
of electroweak processes as well as the electroweak corrections to
some QCD reactions have been calculated. Comparison with the precision
data of LEP, SLC, Tevatron and HERA allows us to test the SM with high
accuracy. In the following subsection we give a short review of the
electroweak 1-loop corrections, essentially following the treatments
of Refs.~\refcite{hollik1,Heinemeyer:2004gx}.

\subsection{Electroweak Radiative Corrections}
The Lagrangian of the SM follows from the construction principles for
gauge theories.
It consists of the gauge field part, the
fermion kinetic terms, the 
gauge interaction terms of the fermion fields, the kinetic and
potential terms of the Higgs doublet, the gauge interaction of
the Higgs doublet, and the terms for the Yukawa interaction
between the fermion and the Higgs field. Their explicit form
will not be given here, but can be found, \mbox{e. g.}, 
in Ref.~\refcite{nachtmann}.

%After spontaneous breaking of the electroweak symmetry, 
The Higgs sector of the SM, after spontaneous symmetry breaking, gets
the following shape:
the Higgs fields $H^{+}$, $H^{+*}$ and $\im H^{0}$ become the
longitudinal components of $W^{\pm}$ and $Z^{0}$. After the shift
$\re H^{0}(x) = \frac{1}{\sqrt{2}}(v+h(x))$ the real scalar
field $h(x)$ becomes the physical Higgs field. Its Lagrangian can
be brought into the form\cite{hunter}
\be
\La_{\mathrm{Higgs}} 
&=&
  \sfrac{1}{2} (\partial_{\mu} h) (\partial^{\mu} h)
- \sfrac{1}{2} m_{h}^{2} h^{2} 
    \left[ 1 + \frac{h}{v} + \frac{1}{4} \left(\frac{h}{v}\right)^{2} 
    \right]
- \sum_{f} \frac{m_{f}}{v} \bar{f} f \, h 
\nn & &
+ \left( 2 \frac{h}{v} + \frac{h^{2}}{v^{2}} \right)
  \left[ m_{W}^{2} W^{+}_{\mu} W^{-\mu} 
       + \sfrac{1}{2} m_{Z}^{2} Z_{\mu} Z^{\mu} \right]
\enspace ,
\label{higgslagrangian}
\ee
where the physical Higgs boson mass at tree level is 
$m_{h}^{2} = 2 \lambda v^{2}$, with
$\lambda$ being the quartic 
coupling constant in the original Higgs potential. 
Eq. (\ref{higgslagrangian}) determines all properties of the 
SM Higgs boson. It has cubic
and quartic self--interactions whose strengths are proportional
to $m_{h}^{2}$. Its couplings to the vector bosons $W^{\pm}$, $Z^{0}$,
and to fermions $f$ are proportional to $m_{W}^{2}$, $m_{Z}^2$,
and $m_{f}$,
respectively. Therefore, the Higgs boson couples dominantly to
the heavy particles. The Higgs boson mass $m_{h}$ is
experimentally not known. In the analyses it is usually treated
as a free parameter of the SM.

The weak vector bosons $W^{\pm}$ and $Z^{0}$ get masses
by the Higgs mechanism,
which are
\be
m_{W}^{2} = \frac{1}{4} g^{2} v^{2}
\enspace , \qquad
m_{Z}^{2} = \frac{1}{4} ( g^{2} + g'^{2} ) v^{2}
          = \frac{m_{W}^{2}}{\cos^{2} \theta_W}
\enspace , 
\label{bosonMasses}
\ee
where $g$ and $g'$ are the $SU(2)$ and $U(1)$ coupling
constants, $\theta_W$ is the electroweak mixing or Weinberg angle, 
$g'/g=\tan\theta_W$, and $v$ is the vacuum expectation value (vev) 
of the $H^{0}$ component of the Higgs field. The photon and $Z^{0}$ 
are linear combinations of the neutral $SU(2)$ and $U(1)$ vector 
bosons with mixing angle $\theta_W$, and the electromagnetic 
coupling is $e = g \sin \theta_W$. Comparison with the muon decay
$\mu^{+} \rightarrow e^{+} \nu_{e} \bar{\nu}_{\mu}$ leads to 
the relation
\be
\frac{1}{\sqrt{2}} G_{\mu} = \frac{g^{2}}{8 m_{W}^{2}} 
= \frac{e^{2}}{8 m_{W}^{2} \sin^{2}\theta_W} 
\enspace , 
\label{FermiConstant}
\ee
where $G_{\mu}$ is the Fermi coupling constant. Inserting the
experimental values\cite{giacommelliTalk,pdg}
\be
G_{\mu} & = & (1.16637 \pm 0.00001) \times 10^{-5} \mbox{ GeV}^{-2}
\enspace ,\label{FermiConstantValues1} \\
\sin^{2}\theta_W & = & 0.23149 \pm 0.00015
\label{FermiConstantValues2}
\ee
together with the fine structure constant
$\alpha = \frac{e^2}{4\pi} = 1/137.03599911(46)$ into Eqs.~(\ref{bosonMasses}) and 
(\ref{FermiConstant}) gives $m_{W} \approx 77.5$~GeV, 
$m_{Z} \approx 88.4$~GeV, and $v \approx 246$~GeV. 
These results for the vector boson masses are already very
close to their experimental values, and historically this was
one of the first triumphs of the SM.
However, when compared with the recent 
experimental values with very small errors\cite{giacommelliTalk,pdg},
\be
m_{W} = 80.425 \pm 0.038\mbox{ GeV}
\enspace , \qquad
m_{Z} = 91.1876 \pm 0.0021\mbox{ GeV}
\enspace , 
\label{bosonMassValues}
\ee
the theoretical values disagree by several standard deviations. 
This shows that the tree--level relations Eqs. (\ref{bosonMasses}) 
and (\ref{FermiConstant}) are not
accurate enough, and that the electroweak loop--corrections have
to be taken into account.

The high precision experiments at LEP, SLAC, and Tevatron have
measured some of the electroweak observables with a very high
accuracy.\cite{giacommelliTalk,pdg,unknown:2003ih}
%, see Fig.~\ref{fig:lepelww}.
For example, the $Z^{0}$ mass 
is known to $0.002\%$, the $W^{\pm}$ mass, the $Z^{0}$ width 
$\Gamma_{Z}$, and some of the partial widths 
$\Gamma (Z^{0} \rightarrow f\bar{f})$, are known to about
$0.1\%$. Some of the forward--backward asymmetries $A_{FB}$ 
and left--right polarisation asymmetries $A_{LR}$ for
$e^+e^- \rightarrow f\bar{f}$ are also measured with very high 
experimental accuracy. In comparison, the electroweak 
radiative corrections are usually of the order of $1\%$, with a
numerical accuracy of about $0.1\%$. This means that we can 
only get a theoretical accuracy comparable to the experimental
one by taking into account the electroweak radiative
corrections. 

%\begin{figure}[t]
%\setlength{\unitlength}{1cm}
%\centering
%\begin{picture}(7.5,9)
%\put(0,0){\framebox(7.5,9){}}
%\put(0.25,-1.5){\epsfig{figure=s04_show_pull_18.eps,scale=0.35}}
%\end{picture}
%\caption{\label{fig:lepelww}Summary of electroweak precision
%measurements, the corresponding fit in the SM and the deviations
%between the two values. From Ref.~\protect\refcite{lepelwwWWW}.}
%\end{figure}

The analysis of the electroweak radiative
corrections provides very accurate tests of the SM and leads to
substantial restrictions on the allowed range of the 
Higgs boson mass.
The $\mathcal{O} (\alpha)$ electroweak corrections at 1--loop level
arise from self--energy diagrams, vertex corrections and box 
diagrams. They affect the basic SM parameters in characteristic 
ways. The self--energy diagrams of the vector bosons play a 
special role. The vacuum polarisation diagrams with charged 
lepton pairs and light quark pairs in the loops lead to a 
logarithmic $q^{2}$ dependence of the electromagnetic coupling.
The bulk of the 1--loop corrections can be taken into account by
including this $q^{2}$--dependence in an effective $\alpha (q^{2})$.
At $q^{2} = m_{Z}^{2}$ this gives
$\alpha (m_{Z}^{2}) = 1/(128.939\pm 0.024)$, where the error is 
mainly due to the uncertainty in the hadronic contribution to 
the vacuum polarisation.\cite{Jegerlehner:2003rx}

The self--energy diagrams of the vector bosons $W^{\pm}$ and $Z^{0}$ 
lead to shifts of their renormalised masses
$m_{Z}^{2} \rightarrow m_{Z}^{2} + \delta m_{Z}^{2}$, and
$m_{W}^{2} \rightarrow m_{W}^{2} + \delta m_{W}^{2}$.
At tree--level we have the relation 
$\sin^{2} \theta_W^{(0)} = 1 - m_{W}^{2}/m_{Z}^{2}$.
At higher orders it is useful to define the effective electroweak mixing angle
\be
\sin^{2} \theta_W 
= \frac{1}{4 |q_f|} \left( 1 - \frac{g_{Vf}}{g_{Af}} \right),
\label{sin2tw}
\ee
where $q_f$ is the electric charge and $g_{Vf}/g_{Af}$ the ratio of
the vector and the axial vector couplings of the fermion $f$.

The parameter $\rho$ is introduced
for comparing the SM predictions with the weak charged and
neutral current data. It is
defined as the ratio between the neutral and charged current
amplitudes. In the SM at tree--level 
$\rho = m_{W}^{2} / ( m_{Z} \cos\theta_W )^{2} = 1$. 
If higher order corrections are taken into
account, or in modifications of the SM, we may have $\rho \ne 1$.
The deviation from $1$, $\Delta \rho $ is a measure of the
influence of heavy particles.
$\Delta \rho$ can be expressed in terms of the vector boson
self--energy contributions.
The main SM 1--loop contribution was calculated in
Ref.~\refcite{veltman} and is
\be
\Delta \rho 
\simeq \frac{3 G_{\mu} m_{t}^{2}}{8 \pi^{2} \sqrt{2}} 
     + \mathcal{O}\left(\frac{m_{b}^{2}}{m_{t}^{2}}\right),
\label{deltarho1}
\ee
where $m_t$ and $m_b$ are the top and bottom quark masses. 
Some of the 2--loop corrections to $\Delta \rho$ have also been
calculated.\cite{hollik1,Heinemeyer:2004gx} They can be of the order of $10\%$ 
of the 1--loop contribution Eq. (\ref{deltarho1}). Experimentally 
we have\cite{Altarelli:2004mr}
$\Delta \rho = (5.4 \pm 1.0) \times 10^{-3}$. 
As can be seen from Eq. (\ref{deltarho1}), the main contribution to 
$\Delta \rho$ comes from heavy particle loops. 

The analysis of the decay $\mu \rightarrow e \nu_{e} \nu_{\mu}$
leads to a relation between $m_{Z}$, $m_{W}$, and the Fermi coupling
constant $G_{\mu}$, which at tree--level is given in 
Eq. (\ref{FermiConstant}). This relation is modified when the 
electroweak radiative corrections to 
$\mu \rightarrow e \nu_{e} \nu_{\mu}$ are taken into account: 
\be
m_{W}^{2} \left( 1 - \frac{m_{W}^{2}}{m_{Z}^{2}} \right) 
= \frac{\pi \alpha}{\sqrt{2} \, G_{\mu}} \cdot \frac{1}{1 - \Delta r} 
\label{mWmasscorrection}
\ee
with
\be
\Delta r 
= \frac{1}{2} \frac{\Delta \alpha}{\alpha} 
- \frac{m_{W}^{2}}{m_{Z}^{2} (1-\frac{m_{W}^{2}}{m_{Z}^{2}})} 
  \Delta \rho \,.
\label{delta-r}
\ee
This shows that the main part of the radiative corrections 
to the tree-level relation (\ref{FermiConstant}) is contained
in the quantity $\Delta r$. There are additional
contributions from vertex corrections and box diagrams.
Numerically $\Delta r$ is in the range $0.04$ to $0.07$. Analogous
to $\Delta \rho$, the main contributions to $\Delta r$
come from heavy particle loops. 

\begin{figure}[t]
\centerline{\epsfig{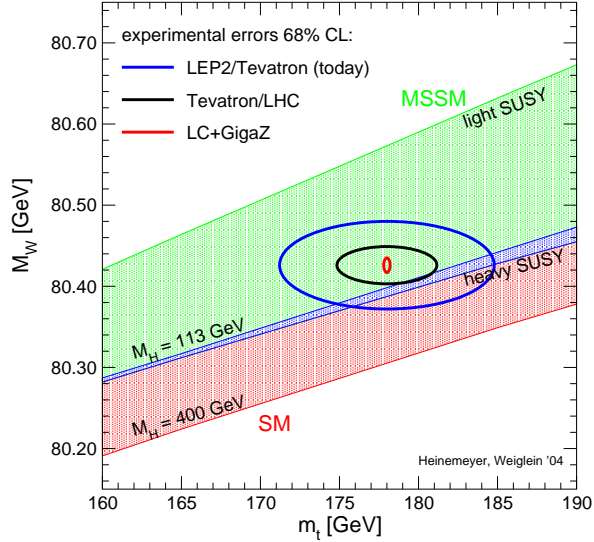}}
\caption{\label{fig:heiwei}
The present experimental accuracy for $m_W$ and $m_t$
after the experiments at LEP and Tevatron
(large ellipse) % (blue)
and the expected accuracies at Tevatron + LHC
(medium size ellipse) % (black)
and LC + GigaZ
(small ellipse). % (red).
The
upper and lower % red and green
bands show the predictions of SM and MSSM,
respectively, where the
small intermediate band % blue band
denotes the overlap between the
predictions of SM and MSSM.
From Ref.~\protect\refcite{HeinemeyerPlot}.}
\end{figure}

In the analysis of the precision data of LEP one usually
proceeds in the following
way:\cite{hollik1,Heinemeyer:2004gx,Altarelli:2004mr} 
The SM parameters $\alpha$, $G_{\mu}$, $m_{Z}$, (see Eqs. 
(\ref{FermiConstantValues1}), (\ref{FermiConstantValues2})
and (\ref{bosonMassValues})),
%C7
the strong coupling 
$\alpha_{s} (m_{Z}) = 0.1187 \pm 0.0020$, 
%Cb$m_{t} = 173.5 \pm 5.2$~GeV,
and $m_{h}$ are taken as the main input 
parameters, and the quantities $m_{W}$ and $\sin^{2} \theta_W$ 
are calculated with the help of Eqs. (\ref{sin2tw}) to 
(\ref{delta-r}). Also the other
$Z$--boson observables are calculated including the electroweak
radiative corrections.
The Higgs boson mass $m_{h}$ is
not known and it is taken as a free parameter
and varied in the allowed range $m_{h} < 1$~TeV. In general, very
good agreement between theory and experiment is obtained.
This can also be illustrated in Fig.~\ref{fig:heiwei}
from Ref.~\refcite{HeinemeyerPlot}, where
the theoretical relation
between the $W^{\pm}$ mass $m_W$ and the
top quark mass $m_t$ in the SM (lower band)
together with the experimental error ellipses from LEP/Tevatron,
Tevatron/LHC and the GigaZ option of an $e^+ e^-$ linear collider
are shown.
This theoretical relation between $m_W$ and $m_t$ is due to the radiative
corrections to the $W^{\pm}$ boson mass, where the loops involving the top
quark play a special role. The leading corrections depend quadratically on
$m_t$ and logarithmically on the Higgs boson mass $m_h$.
While in this calculation essentially all basic
electroweak parameters enter, $m_W$ depends very significantly on $m_t$
and on $m_h$.
The width of the SM band is mainly due to the variation of the Higgs boson
mass in the range $113$~GeV $\lesssim m_h \lesssim 400$~GeV.
If the Higgs boson mass is 
left as a free parameter and a global fit to the precision data 
is performed, the best fit is obtained for the value
$m_{h} = 114^{+69}_{-45}$~GeV, or equivalently, $m_{h} < 260$~GeV
at $95\%$ confidence level.\cite{lepelwwWWW}
This result is consistent with the
present experimental lower bound from LEP2,
$m_{h} > 114.4$~GeV.\cite{giacommelliTalk,higgsbound}

From this analysis we learn that a heavy particle can be probed 
when it appears in the virtual
loops of the quantum corrections. As these loop effects
influence some of the measurable observables, it may be
possible to derive limits on the allowed mass range of this heavy
particle. This is possible 
%C8
although the energy is not high enough
to directly produce the heavy particle in experiment. The prize
to be paid is, of course, high precision in experiment as well
as in the theoretical calculation.

\subsection{Higgs Boson Searches}

For a complete verification of the SM
and its electroweak symmetry breaking mechanism
we have to find the Higgs boson.
In the preceding section we have seen that consistency of the SM with the
existing precision data requires 
$m_{h} < 260$~GeV.
Combining the final results from LEP2 of the four experiments
ALEPH, DEPLPHI, L3 and OPAL a lower bound for the SM
Higgs boson mass of $114.4$~GeV at $95\%$ confidence level
arises.\cite{giacommelliTalk,higgsbound}
A considerable part of the allowed mass range for the Higgs boson
is within the reach of Tevatron. A full coverage of this mass range
will be provided by LHC and a future $e^{+}e^{-}$ linear collider or
muon collider. The search for the Higgs boson, therefore, has
high priority at all present and future colliders. In this
section we will discuss the principle ideas of Higgs boson
searches at the Tevatron, LHC and a future $e^{+}e^{-}$ linear collider.

The main production mechanisms at hadron colliders are
gluon--gluon fusion, $WW$ or $ZZ$ fusion,
associated production with $W$ or $Z$ and
associated production with $t\bar{t}$ or $b\bar{b}$.\cite{Djouadi:2002pw}
At the $p\bar{p}$ collider Tevatron with $\sqrt{s}=2$~TeV
the most relevant production mechanism is the
associated production with $W$ or $Z$ bosons, where a detectable rate
of Higgs events is expected for $m_h=120$~GeV and an integrated luminosity
$\int\mathcal{L} = 2~\mathrm{fb}^{-1}$.
For example, a clear signature is expected for the reaction\cite{tevhiggs}
\be
p + \bar{p} \rightarrow W^{\pm} + h \rightarrow 
\ell^{\pm} + (q\bar{q}) + p_{\mathit{T}\mathrm{miss}} 
\label{higgsstrahlung}
\ee
where $85\%$ of the $q\bar{q}$ pairs are $b\bar{b}$, and the 
$\nu_{\ell}$ from 
$W^{\pm} \rightarrow \ell^{\pm} \nu_{\ell}, \ell=e, \mu$
is reconstructed from the missing transverse momentum 
$p_{\mathit{T}\mathrm{miss}}$.
The $WW$ or $ZZ$ fusion cross sections are slightly smaller for $m_h
\lesssim 150$~GeV. The cross sections for associated production
with $t\bar{t}$ or $b\bar{b}$ are rather low.

The dominant production mechanism at the LHC with $\sqrt{s}=14$~TeV
is gluon--gluon fusion with a cross section
$\gtrsim 10~\mathrm{pb}$ for $m_h < 260$~GeV. The cross section for
$WW$ or $ZZ$ fusion is of the order of a few pb, whereas the cross
sections for the associated productions with gauge bosons or 
$t\bar{t}$, $b\bar{b}$
may contribute for lower Higgs masses.

\begin{figure}[t]
\centerline{\epsfig{file=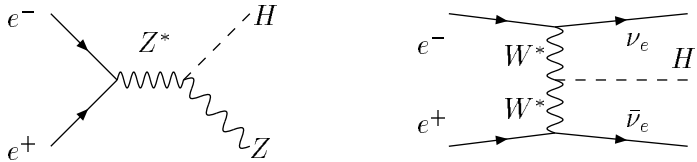,scale=1,clip}}
\caption{\label{fig:higgsfeyndiag}
Feynman diagrams for the Higgsstrahlung and $WW$ fusion production
mechanisms of SM Higgs bosons in $e^+ e^+$ annihilation.}
\end{figure}

The search for the Higgs boson will also have a very high priority at
a future linear collider.\cite{lincoll}
The production of a SM Higgs boson
in $e^+ e^-$ annihilation can proceed via ``Higgsstrahlung''
$e^+ e^- \to Z h$, $WW$ fusion $e^+ e^- \to \nu_e \bar{\nu}_e h$, and
$ZZ$ fusion $e^+ e^- \to e^+ e^- h$.
%The relative importance of Higgsstrahlung and $WW$ fusion is shown in
%figure~\ref{fig:SMHiggs} (from \cite{LC2,Djouadi:2002pw}).
At $\sqrt{s} = 500$~GeV the Higgsstrahlung process dominates for
$m_h \gtrsim 160$~GeV, whereas for $m_H \lesssim 160$~GeV the $WW$
fusion process gives the largest contribution. The higher $\sqrt{s}$
the more important is the $WW$ fusion process. 
The Feynman diagrams for the Higgsstrahlung and the $WW$ fusion
processes are shown in Fig.~\ref{fig:higgsfeyndiag}.
Only for
$\sqrt{s} \gtrsim 800$~GeV the $ZZ$ fusion process can contribute
about $\gtrsim 10\,\%$ of the total production rate.
%As can also be seen in figure~\ref{fig:SMHiggs}, if
If $m_h \lesssim 260$~GeV as suggested by the electroweak precision data,
an optimal choice for the c.m.s.\ energy is $\sqrt{s} \approx 350$ --
$500$~GeV.

In conclusion one can say that there are some prospects 
%C9
of finding the Higgs boson at the Tevatron. The LHC will cover the
full mass range up to $m_{h} \approx 1$~TeV. Precise
determinations of all important Higgs boson couplings will be
possible at a future $e^{+}e^{-}$ linear collider or muon
collider.\cite{lincoll,tevhiggs}

\section{Grand Unification}

We can study the scale dependence of the three gauge coupling
``constants'' with the help of the renormalization group
equations (RGE). If we evolve 
the strong, electromagnetic and weak coupling constants to
higher energy scales, they become
approximately equal at $M_{U} \approx 10^{14}$~GeV to
$10^{16}$~GeV, the grand unification scale. 
This behaviour of the gauge coupling constants suggests that
the SM is embedded into an underlying grand unified 
theory (GUT). If we assume that this GUT is also a gauge theory,
its symmetry group has to be semi-simple and it has to contain 
$SU(3) \times SU(2) \times U(1)$ 
as a subgroup.  The GUT gauge group is unbroken 
at energies higher than the GUT scale $M_{U}$, and is
spontaneously broken to the SM gauge group at lower energies.
The smallest semi-simple GUT group with rank 4 is $SU(5)$. 
Other possible choices are $SO(10)$, $E(6)$ etc.
In this section we will shortly mention the basic features of 
$SU(5)$ and $SO(10)$ grand unification
(for other examples see Refs.~\refcite{ross,moha,Mohapatra:1999vv}). 

In the $SU(5)$ GUT model the 15 helicity states of each family of
quarks and leptons are put into the $\mathbf{\bar{5}}$ 
($e_{L}$, $\nu_{eL}$, $d^{C}_{iL}$, $i=1,2,3$) and
$\mathbf{10}$
($e^{C}_{L}$, $u_{iL}$, $u^C_{iL}$, $d_{iL}$, $i=1,2,3$)
representations.
Here the right--handed states 
$f_{R}$ are written as the charge conjugate left--handed states 
$f^{C}_{L}$, and $i=1,2,3$ denotes the three colours of the quarks. 
Furthermore the $SU(5)$ GUT model contains 24 vector bosons
corresponding to the 24 generators of the Lie group $SU(5)$,
i.e.\ the gluons, the electroweak gauge bosons and 12 new coloured and
charged gauge bosons called $X$ and $Y$ which are leptoquarks and diquarks.
The spontaneous breaking of $SU(5)$ can be achieved in a two-step
procedure. In a first step a $\mathbf{24}$ multiplet of scalar Higgs
fields with masses $\mathcal{O}(M_{U})$ breaks $SU(5)$ to the SM group
$SU(3) \times SU(2) \times U(1)$. In a second step the SM group is
broken to $SU(3) \times U(1)$ by a $\mathbf{5}$ multiplet of Higgs
fields with masses $\mathcal{O}(m_{Z})$.

The gauge bosons $X$ have couplings of the form $X \ell q$ with
leptons and quarks and
can therefore induce proton decay, 
for example,  $p \rightarrow \pi^{0} e^{+}$.
The mass of the vector bosons $X$ has to be of
the order $m_{X} \approx M_{U}$. 
The order of
magnitude for the proton lifetime can be estimated as 
$\tau_{p}^{-1} \approx \alpha_{U}^{2} m_{p}^{2}/m_{X}^{2}$, 
where $m_{p}$ is the proton mass. 
In the non-supersymmetric $SU(5)$ GUT model 
with $M_{U} \approx 10^{14}$~GeV
one obtains
$\tau_{p} \approx 10^{30}$~years,
whereas the present experimental lower
bound for the proton lifetime is $\tau_{p} > 1.9 \times
10^{33}$~years.
In the supersymmetric $SU(5)$ GUT model the
unification scale turns out to 
be $M_{U} \approx 10^{16}$~GeV. This leads to a larger value for the
proton lifetime, which is in agreement with the experimental
lower bound although the parameter space of the supersymmetric $SU(5)$
GUT is tightly constrained.\cite{Bajc:2002bv}

The supersymmetric $SO(10)$ GUT model has a number of additional
desirable features compared to $SU(5)$.\cite{Mohapatra:1999vv,Baer:2000jj}
For example, the 15 helicity states of
quarks and leptons together with a SM gauge singlet right-handed
neutrino state leading to nonzero neutrino masses
are included in one $\mathbf{16}$
representation of $SO(10)$.
Furthermore the $SO(10)$ GUT model can solve the SUSY CP and
$R$-parity problems because it is left-right symmetric.
There are many ways to break $SO(10)$ down to the SM, details can be
found in Ref.~\refcite{Mohapatra:1999vv}.

\section{Supersymmetry}
Supersymmetry (SUSY) is a new symmetry relating bosons and fermions.
The particles combined in a SUSY multiplet have
spins which differ by $\frac{1}{2}$.
This is different from the symmetries of the SM or a GUT
where all particles in a multiplet have the same spin (for an
introduction to SUSY see e.g. Ref.~\refcite{wess}).

SUSY is at present one of the most attractive and best studied
extensions of the SM.
The most important motivation for that is the fact that
SUSY quantum field theories have in general better high--energy
behaviour than non--SUSY ones.
This is due to the cancellation of the divergent
bosonic and fermionic contributions to the 1-loop radiative corrections. 
A particularly important example is the cancellation
of the quadratic divergencies in the loop corrections to the 
Higgs mass. This cancellation mechanism provides one of
the best ways we know to stabilize the mass 
of an elementary scalar Higgs field against radiative
corrections and keep it ``naturally'' of the order 
$\mathcal{O} (m_{Z})$. 

%\subsection{The Minimal Supersymmetric Standard Model}

Practically all SUSY modifications of the SM are based on local $N=1$
SUSY.
In the ``minimal'' SUSY extension of the SM a hypothetical SUSY partner
is introduced for every known SM particle.
The SUSY partners of the neutrinos, leptons, and
quarks are called scalar neutrinos 
$\tilde{\nu}$, left and right scalar
leptons $\tilde{\ell}_{L}$, $\tilde{\ell}_{R}$, and left and right 
scalar quarks $\tilde{q}_{L}$, $\tilde{q}_{R}$, respectively. They
have spin $0$. The SUSY partners of the gauge vector bosons have
spin $\frac{1}{2}$ and are called gauginos. The 
photino $\tilde{\gamma}$, $W^{\pm}$-ino $\tilde{W}^{\pm}$, $Z$-ino 
$\tilde{Z}$, and gluino $\tilde{g}$ are the partners of $\gamma$, 
$W^{\pm}$, $Z^{0}$, and the gluon, respectively. In
the local version of SUSY the graviton gets a spin--$\frac{3}{2}$
SUSY partner, called gravitino. Furthermore, at least two
isodoublets of Higgs fields $H_{i}, i=1,2$, have to be introduced,
together with their SUSY partners, the higgsinos $\tilde{H}_{i},
i=1,2$, which have spin $\frac{1}{2}$. 
In this way the anomalies in the triangular loops cancel.
%The quantum
%numbers assigned to the two Higgs doublets are such that $H_{1}^{0}$
%gives mass terms for the charged leptons and down--type quarks,
%and $H_{2}^{0}$ to the up--type quarks. 
%The gauge couplings and Yukawa couplings of the SUSY partners are 
%the same as those of the corresponding SM particles.
The model obtained in
this way is the Minimal Supersymmetric Standard Model (MSSM).\cite{nilles}
%The total Lagrangian contains the SM part, and its SUSY 
%counterpart. In addition there is a mixing term of the 
%Higgs fields $H_{1}$ and $H_{2}$ with mass parameter $\mu$, together 
%with the corresponding mixing term of the Higgsino fields 
%$\tilde{H}_{1}$ and $\tilde{H}_{2}$.
In the ``next-to-minimal'' SUSY extension of the SM (NMSSM) an
additional Higgs singlet and the corresponding higgsino are
introduced (see for example Ref.~\refcite{NMSSM} and References therein).

The gauginos and higgsinos form quantum
mechanically mixed states. The charged and neutral mass
eigenstates are the charginos $\tilde{\chi}_{i}^{\pm}$, $i=1,2$, and
neutralinos $\tilde{\chi}_{i}^{0}$, $i=1,\dots 4$, respectively.
The left and right 
states of the scalar fermions are also mixed, with a mixing term
proportional to the corresponding fermion mass. Therefore, the
mass eigenstates of the first and second generation scalar
fermions are to a good approximation the left and right states.
However, there may be strong left--right mixing in the sector 
of the scalar top and bottom quarks and the scalar tau lepton.

\begin{figure}[t]
\setlength{\unitlength}{1mm}
\begin{center}
\begin{picture}(56,60)%113,60
%\put(0,0){\framebox(56,60)}
\put(-1,-57){\mbox{\epsfig{figure=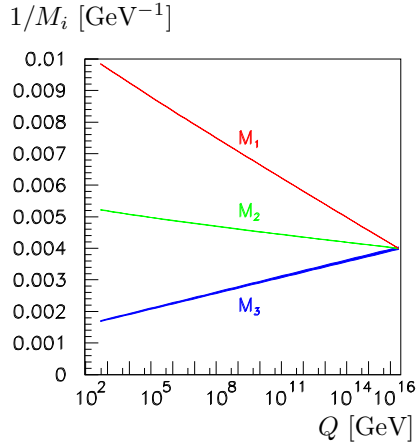,height=12cm}}}
%\put(47,47){(a)}
\put(1,55.5){\mbox{$1/M_i$~[GeV$^{-1}$]}}
\put(42,1){\mbox{$Q$~[GeV]}}
%\put(56,-57){\mbox{\epsfig{figure=gen1_lhclc.eps,height=12cm}}}
%\put(104,47){(b)}
%\put(61,55.5){\mbox{$M^2_{\tilde f}$~[$10^3$ GeV$^2$]}}
%\put(100,1){\mbox{$Q$~[GeV]}}
\end{picture}
\end{center}
\caption{Evolution of the gaugino mass parameters $M_i$
from low to high scales.
From Ref.~\protect\refcite{Allanach:2004ed}.}
\label{fig:runningMi}
\end{figure}

If SUSY was an exact symmetry, then the masses 
of the SUSY partners would be the same as those of
the corresponding SM particles.
This is evidently not observed in nature, therefore,
SUSY must be broken. Essentially, the idea is to
break local SUSY spontaneously at a high 
energy scale.\cite{Arnowitt}
The result is the 
global SUSY Lagrangian plus 
additional ``soft SUSY--breaking terms'', which are
mass terms for the SUSY partners, and additional 
trilinear coupling terms for the scalar fields.\cite{ross,moha,nilles}
%These additional terms in the 
%Lagrangian are ``soft--breaking terms'', because they lead 
%at most to logarithmic divergences. However, as we do not know the 
%SUSY breaking mechanism we have to make
Further assumptions are necessary to fix the 
additional soft--breaking parameters.
For example, we can
assume that at the GUT scale $M_{U}$ all scalar SUSY 
partners have the same mass $M_{0}$, all gauginos have a common 
mass $M_{1/2}$, and all trilinear couplings of the scalar fields 
have a common strength $A_{0}$. 
We obtain their values at the weak scale by evolving them
with the RGEs from $Q = M_{U}$
to $Q \approx M_Z$.\cite{Blair:2002pg}
The model obtained in this way is called 
constrained MSSM (CMSSM) or minimal supergravity-inspired model 
(mSUGRA). In Fig.~\ref{fig:runningMi} we show an example where we plot the
gaugino mass parameters $M_{1}$, $M_{2}$, $M_{3}$
%the mass parameters $M_{\tilde{f}}^2$
%$D_1^2$, $Q_1^2$, $U_1^2$, $E_1^2$, $L_1^2$
%of the first/second generations scalar fermions 
%and the Higgs mass parameter $M_{H_2}^2$
as a function of the scale $Q$. 

Radiative electroweak symmetry breaking is
a further attractive feature of SUSY.
%The masses of the two neutral Higgs fields 
%$H_{1}^{0}$ and $H_{2}^{0}$ depend logarithmically on the 
%energy scale.
This can be achieved by exploiting the logarithmic scale dependence of
the squares of the masses of the Higgs fields $H_{1}^{0}$ and $H_{2}^{0}$.
Starting at the scale $Q = M_{U}$
with the mass values $M_{H_{1}}^{2} = M_{H_{2}}^{2} = M_{0}^{2}$ 
and evolving to lower energies,
it turns out that $M_{H_{2}}^{2}$ can become negative at 
$Q \approx M_Z$.
The reason is that $M_{H_{2}}^{2}$ gets large negative 
contributions from the top--quark loops. In this way 
spontaneous breaking of the electroweak symmetry is induced. 
$H_{1}^{0}$ and $H_{2}^{0}$ get vev's  
$\langle H_{i}^{0} \rangle = \frac{1}{\sqrt{2}} v_{i},i=1,2$, 
and the vector bosons get masses 
$m_{W}^{2}=\frac{1}{4} g^{2} (v_{1}^{2}+v_{2}^{2})$ and
$m_{Z}^{2}=\frac{1}{4} (g^{2}+g'^{2}) (v_{1}^{2}+v_{2}^{2})$. 
%In this scheme the 
%scale of electroweak symmetry breaking is determined by the
%parameters specified at the scale $M_{U}$. Even if SUSY is
%broken, there is a partial cancellation between the bosonic and
%fermionic loop corrections to the Higgs boson mass. As long as
%the mass difference between the SM particles and their SUSY
%partners is less than approximately $1$~TeV, the resulting Higgs
%boson mass is of the order of the weak scale without fine tuning
%of the parameters.
%The ratio 
%$\tan\beta=v_{2}/v_{1}$ is not determined and is a free parameter
%of the MSSM.  
This mechanism works because the top--quark mass is much larger than
the other quark masses (as one can show $m_t > 60$~GeV must be
fulfilled). Furthermore, the mass difference between the SM particles
and their SUSY partners must be less than about 1~TeV.

%As two Higgs doublets have to be introduced
The Higgs sector of the MSSM
%is more complicated than that of the SM. After spontaneous electroweak
%symmetry breaking there remain 
contains five Higgs bosons, the $CP$--even $h^{0}$ and $H^{0}$, the
$CP$--odd $A^{0}$, and a pair of charged ones, $H^{\pm}$.\cite{guha} An
important prediction of the MSSM is that the mass of the lighter 
$CP$--even state $h^{0}$ is always $m_{h^{0}} < m_{Z}$ at 
tree--level. There are large radiative corrections which change
this prediction to 
$m_{h^{0}} \lesssim 140$~GeV.\cite{Carena:2002es,Heinemeyer:2004ms}
%A significant part of this mass range lies within the reach of Tevatron.
Comparing with the discussion in subsection 2.1 we see
that this prediction for $m_{h^{0}}$ lies within the allowed range
%almost coincides with the most probable value 
for the Higgs boson mass
obtained in the analysis of the electroweak precision data.
We note in passing that some SUSY parameters may be complex and induce
CP-violating effects, for example, mixing between the CP-odd $A^0$ and
the CP-even $h^0$ and $H^0$.\cite{Heinemeyer:2004ms,Carena:2000yi}

\begin{figure}[t]
\setlength{\unitlength}{1mm}
\begin{center}
\begin{picture}(113,57)
%\put(0,0){\framebox(113,57)}
\put(-15,-68){\mbox{\epsfig{figure=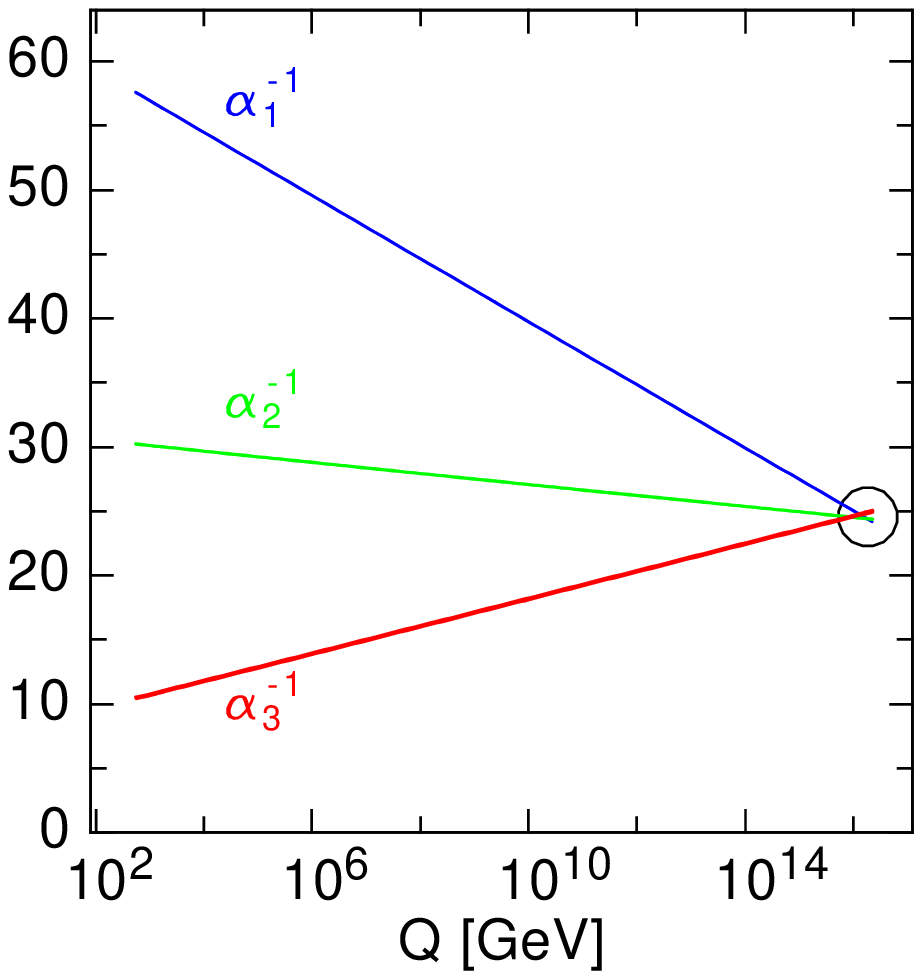, height=15.5cm}}}
%\put(76,20){\mbox{\huge $\Rightarrow$}}
\put(44,-68){\mbox{\epsfig{figure=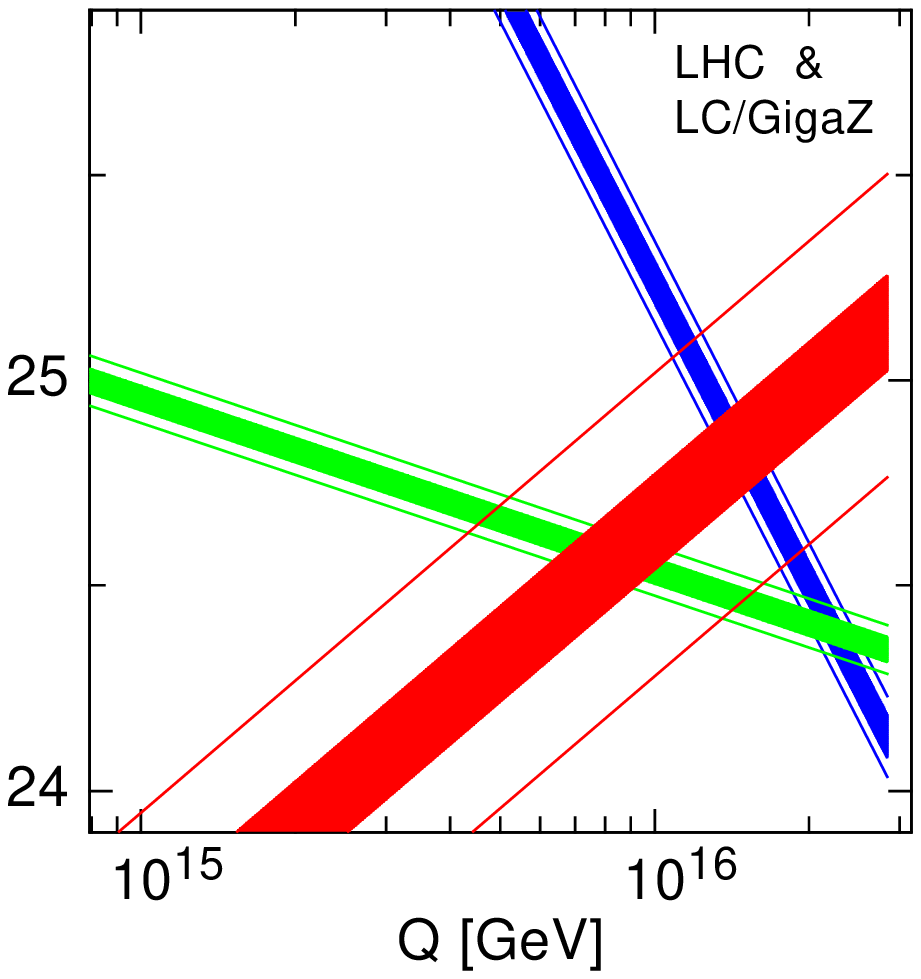, height=15.5cm}}}
\put(45,50){(a)}
\put(67,50){(b)}
\end{picture}
\end{center}
\caption{(a) Running of the inverse gauge couplings from low
  to high energies.
  (b) Expansion of the area around $Q = 10^{16}$~GeV. 
      The wide error bands are based on present data, and the spectrum
      of supersymmetric particles from LHC measurements within mSUGRA.
      The narrow bands demonstrate the improvement expected by future
      GigaZ analyses and the measurement of the complete
      spectrum at ``LHC+LC''. From Ref.~\protect\refcite{Allanach:2004ed}.}
\label{fig:runningalpha}
\end{figure}

It turns out that
the unification of the three gauge couplings works better in the
MSSM than without SUSY.\cite{Blair:2002pg}
We illustrate this in Fig.~\ref{fig:runningalpha}, where we plot
the gauge couplings in the MSSM as a function of the
energy scale. The evolution in the MSSM is different from that
in the SM, because the RGEs of the 
MSSM contain also the contributions from the SUSY
particles. In the MSSM the unification scale turns out to
be of the order $M_{U} \approx 2 \times 10^{16}$~GeV, provided the masses
of the SUSY particles are not much larger than approximately
$1$~TeV.
%Moreover, the prediction for $\sin^{2}\theta_W$ 
%agrees better with the experimental value than in non--SUSY GUT models.

%\subsection{Experimental Search for Supersymmetric Particles} 

The experimental search for SUSY particles has a high priority at
present colliders and will become even more important at LHC and the
future $e^+e^-$ linear collider ILC. In the discussion of the possible
signatures one has to distinguish the two cases whether the
multiplicative quantum number $R$-parity $R_{P}=(-1)^{3B+L-2S}$ is
conserved or violated.
%In the MSSM there is the multiplicative quantum number
%$R$-parity $R_{P}=(-1)^{3B+L-2S}$ which is conserved.
SUSY particles have $R_{P}=-1$ and ordinary particles have $R_{P}=+1$.
%This means that SUSY particles must be produced in pairs, and
%that a SUSY particle decays into ordinary particles and 
%another SUSY particle until the lightest SUSY particle (LSP) 
%is reached. 
If $R_{P}$ is conserved, then there exists a lightest SUSY particle
(LSP) which is stable. 
Cosmological arguments suggest that it is neutral and only
weakly interacting.
It is an excellent candidate for dark matter.
We assume that the lightest neutralino $\tilde{\chi}_{1}^{0}$
is the LSP, which in the CMSSM holds in most of the parameter space.
In experiment the LSP
behaves like a neutrino and its energy and 
momentum are not observable. 
Therefore, in the $R_{P}$ conserving case
the characteristic experimental signatures for SUSY
particles are events with missing energy
$E_{\mathrm{miss}}$ and missing momentum $p_{\mathrm{miss}}$.

In the $R_{P}$ violating case the SUSY Lagrangian contains additional
terms which are allowed by SUSY and the gauge symmetry, but are lepton
number violating and/or baryon number violating. Consequently, the LSP
is not stable and decays into SM particles. Therefore, in the $R_{P}$
violating case the $E_{\mathrm{miss}}$ and $p_{\mathrm{miss}}$
signature is in general not applicable. However, due to the decay of
the LSP there are more leptons and/or jets in the final state. At an
$e^+e^-$ collider the main signature for $R_{P}$ violation is,
therefore, an enhanced rate of multi-lepton and/or multi-jet final
states. If the mean decay length of the LSP is too large and it decays
outside the detector, then its energy and momentum remain invisible
and the $E_{\mathrm{miss}}$ and $p_{\mathrm{miss}}$ signature is again
applicable. If the LSP decays within the detector and the decay length
is long enough, then displaced vertices may occur, which then provide
a further important observable for $R_{P}$ violating SUSY. At a hadron
collider the situation may be more involved. If the lepton number
violating terms dominate over the baryon number violating ones, then
the enhanced number of multi-lepton final states is again a good
signature.

$R_{P}$ violating SUSY can also provide a viable framework for
non-vanishing neutrino masses and a quantitative description of the
present data on neutrino oscillations. This is a very attractive
feature of $R_{P}$ violating SUSY, which in its bilinear formulation
can be shortly described in the following way
(for a review see Ref.~\refcite{Hirsch:2004he}):
As lepton number is not conserved, the neutrinos mix with the
neutralinos and the charged leptons mix with the charginos, where the
amount of mixing depends on the $R_{P}$ violating parameters. In
bilinear $R_{P}$ violating SUSY one neutrino gets a non-vanishing mass
already at tree level, while the other two neutrinos get their masses
at 1-loop level. In this way ``small'' neutrino masses are obtained
and the data on neutrino oscillations can be quantitatively described.
After fixing the $R_{P}$ violating parameters by the solar and
atmospheric neutrino data, the $R_{P}$ violating decay widths of the
SUSY particles can be predicted. This means that the low energy
phenomena in the neutrino sector are linked to the SUSY particle
sector, which we expect to probe at high energy colliders.

At LEP no supersymmetric particles have been found.\cite{LEPSUSYWG}
This implies lower mass bounds which are
$m_{\tilde{\chi}^\pm_1} > 103.5$~GeV (for $m_{\tilde{\nu}_e} > 300$~GeV),
$m_{\tilde{e}} > 99.9$~GeV,
$m_{\tilde{\mu}} > 94.9$~GeV,
$m_{\tilde{\tau}} > 86.6$~GeV,
$m_{\tilde{t}} > 95$~GeV and
$m_{\tilde{b}} > 94$~GeV.
The limit on
the mass of $\tilde{\chi}_{1}^{0}$ is model dependent. Within the 
CMSSM the non--observation of charginos and neutralinos excludes
certain CMSSM parameter regions. From these follows the limit on
the $\tilde{\chi}_{1}^{0}$ mass $m_{\tilde{\chi}_{1}^{0}} > 50.3$~GeV. 
The non--observation of Higgs bosons leads to the mass limits 
$m_{h^{0}} > 92.9$~GeV and $m_{A^{0}} > 93.3$~GeV in the CP-conserving
MSSM with real parameters. In the CP-violating MSSM with complex
parameters no universal lower bound for the masses of the neutral
Higgs bosons can be defined.\cite{LEPHIGGSWG} 

At the Tevatron the strong interaction processes of gluino and
squark production, $p\bar{p} \rightarrow \tilde{g} \tilde{g}$, 
$\tilde{g} \tilde{q}$, $\tilde{q} \bar{\tilde{q}}$, are the SUSY 
reactions with the highest cross sections. Gluinos and squarks 
may have cascade decays which start with 
$\tilde{g} \rightarrow q\bar{q} \tilde{\chi}_i^{0}$, 
$q\bar{q}' \tilde{\chi}_{i}^{\pm}$, 
$\tilde{q} \rightarrow q \tilde{\chi}_{i}^{0}$, 
$\tilde{q} \rightarrow q'\tilde{\chi}_{i}^{\pm}$, and continue until 
the LSP $\tilde{\chi}_{1}^{0}$ is reached. Suitable kinematical cuts 
are necessary to distinguish a possible signal from the huge 
SM background. The present gluino and squark mass limits are 
$m_{\tilde{g}} \gtrsim 400$~GeV, and $m_{\tilde{q}} \gtrsim 250$~GeV if 
$m_{\tilde{q}} \approx m_{\tilde{g}}$, 
$m_{\tilde{q}} \gtrsim 200$~GeV if $m_{\tilde{g}} \approx 500$~GeV
whereas for $m_{\tilde{g}} \gtrsim 560$~GeV
no limit on the squark mass can be obtained from measurements at
Tevatron.\cite{ICHEP04}
%This squark mass limit
%does not hold for the scalar top and bottom quarks which has to be
%treated differently.
For $\tilde{t}_{1}$ and $\tilde{b}_{1}$ the mass limits are
different: 
$m_{\tilde{t}_{1}} \gtrsim 115$~GeV provided
$m_{\tilde{\chi}_{1}^{0}} \lesssim 50$~GeV and
$m_{\tilde{b}_{1}} \gtrsim 140$~GeV provided
$m_{\tilde{\chi}_{1}^{0}} \lesssim 70$~GeV, respectively.\cite{LEPSUSYWG}
Another interesting SUSY reaction which can be studied at the 
hadron colliders is 
$p\bar{p} \rightarrow \tilde{\chi}_{1}^{\pm} \tilde{\chi}_{2}^{0}$. 
It leads to the very clean signature 
$3 \ell + p_{\mathrm{miss}}$, $\ell=e,\mu$.\cite{Baer:1992dc}
The Tevatron mass limit
for $\tilde{\chi}_{1}^{\pm}$,
following from the non-observation of this reaction,
is close to the LEP limit. At the 
upgraded Tevatron the expected SUSY mass reach will be 
$m_{\tilde{g}} \approx m_{\tilde{q}} \approx 390$~GeV, 
$m_{\tilde{t}_{1}} \approx 180$~GeV, 
$m_{\tilde{\chi}_{1}^{\pm}} \approx 250$~GeV, for an integrated 
luminosity of $2~\mathrm{fb}^{-1}$. 

\begin{figure}[t]
\setlength{\unitlength}{1mm}
\begin{center}
\begin{picture}(113,54)
%\put(0,0){\framebox(113,54)}
\put(-1,-2){\epsfig{figure=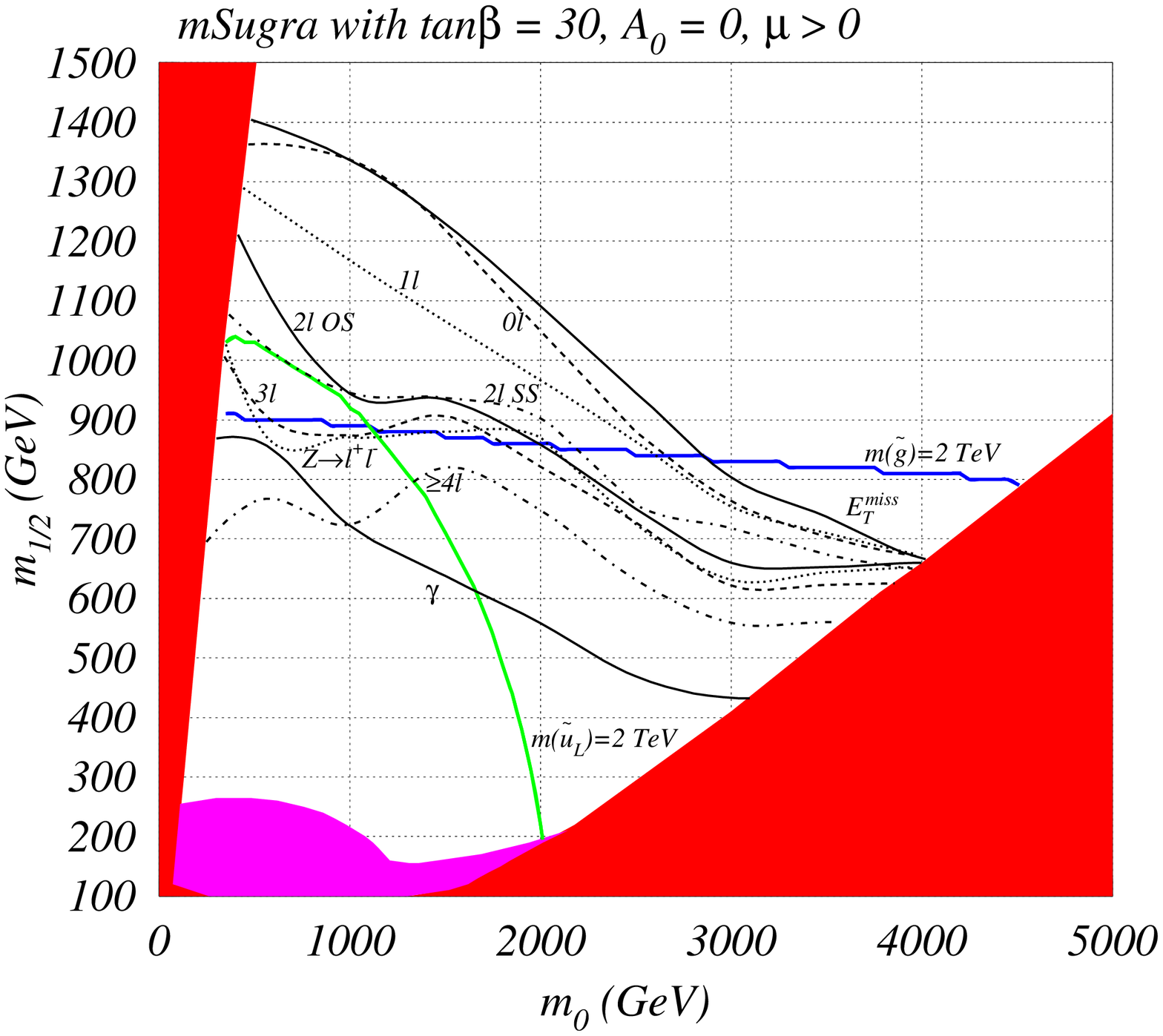,scale=0.3}}
%\put(-1,-2){\epsfig{figure=LHCmSUGRA.eps,scale=0.29}}
\put(51,51){(a)}

\put(54.5,-2){\epsfig{figure=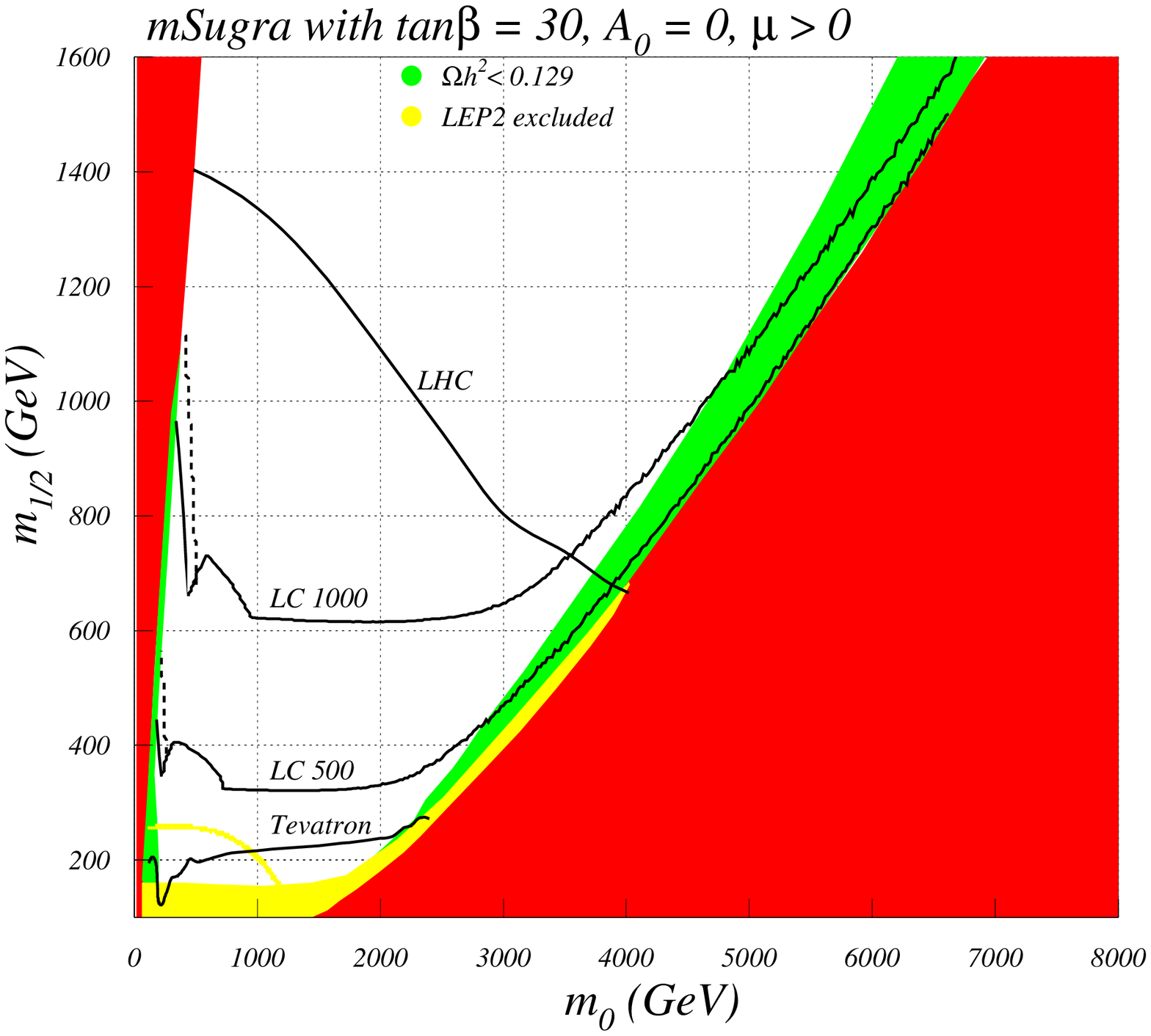,scale=0.3}}
%\put(56.5,-2){\epsfig{figure=LCmSUGRA.eps,scale=0.29}}
\put(106,51){(b)}

\end{picture}
\end{center}
\caption{\label{fig:mSUGRAatColliders}
(a) The reach of the LHC 
for various production channels of SUSY particles
in the the mSUGRA model for
$\tan\beta =30$, $A_0=0$ and $\mu >0$,
assuming 100 fb$^{-1}$ of integrated luminosity.
The shaded regions are excluded by theoretical and experimental
constraints. From Ref.~\protect\refcite{Baer:2003wx}.
(b) Reach of an $e^+ e^-$ linear collider with
$\sqrt{s}=0.5$ and 1~TeV in
the mSUGRA model for $\tan\beta =30$, $A_0=0$ and $\mu >0$.
For comparison the reach of the Tevatron assuming 10 fb$^{-1}$
of integrated luminosity (for isolated trileptons) and the reach of
the LHC (in the ``inclusive'' $\not\!\!E_T$ channel)
assuming 100 fb$^{-1}$ of integrated luminosity is shown.
The dark shaded region is excluded by theoretical constraints.
The gray shaded region shows points where the relic
density $\Omega h^2<0.129$ as preferred by WMAP.
The light shaded region is excluded by LEP2.
From Ref.~\protect\refcite{Baer:2003ru}.
}
\end{figure}

At LHC gluinos and squarks will be detectable up to masses of
approximately $1-2$~TeV, as is illustrated in 
Fig.~\ref{fig:mSUGRAatColliders}. The
%C27
cascade decays of these particles
will play an important role.\cite{lhc} On the one hand they 
will give rise to characteristic signatures, for example the 
same--sign dilepton signature of gluinos. On the other hand, 
in the cascade decays the weakly interacting charginos and 
neutralinos will appear whose properties can also be studied. 
If weak--scale SUSY is not found at the Tevatron, 
then the LHC is the collider where it will be either discovered 
or definitely disproved. 

The reach in the mSUGRA parameter space
of an $e^+ e^-$ linear collider with $\sqrt{s} = 0.5$ to
1~TeV will be be somewhat smaller than that of the LHC
(Fig.~\ref{fig:mSUGRAatColliders} (b)). 
However, due to the high luminosity and good energy resolution 
expected an $e^+ e^-$ linear collider will be inevitable for
precision measurements, especially in the neutralino and chargino
sectors.
This will enable us to determine very precisely the SUSY parameters
and to reconstruct the underlying
theory.\cite{lincoll,Blair:2002pg,LHCLCreport}
However, the signatures will
be more complicated than those at LEP, because also the heavier SUSY
particles will be produced which have cascade decays. This
will lead to characteristic events with several leptons and/or
jets, and missing energy and momentum. 

Inspecting again Fig.~\ref{fig:heiwei}
it can be seen that already the precision data obtained at the GigaZ
mode of the linear collider (small ellipse) will presumably allow us
to discriminate between the 
SM and the MSSM or another extension of the SM.
The present experimental errors
(large ellipse)
do not allow to discriminate between the two models.
In this figure the MSSM band is
obtained by varying the SUSY parameters in the range allowed by the
experimental and theoretical constraints. There is a small overlap of
the SM and MSSM bands
(small intermediate band)
for a light Higgs boson ($m_h = 113$~GeV) and a heavy SUSY
spectrum.

\section{Extra dimensions}

A solution to the hierarchy problem can in principle be obtained by
formulating gravity in $4 + \delta$ dimensions, where
$\delta = 1,2,3,\ldots$ are the so-called 
``extra'' dimensions,\cite{add,Randall:1999ee}
which are assumed to be compactified  with a radius $R$
(for a review see Ref.~\refcite{Rizzo:2004kr}).
In the model of Ref.~\refcite{add} it is assumed that SM physics is restricted
to the 4-dimensional brane, whereas gravity acts in the $4 + \delta$
dimensional bulk.
In 4-dimensional space-time the Planck mass is
$M_{\mathrm{Pl}} = 1.2 \cdot 10^{19}$~GeV. In the
$(4 + \delta)$-dimensional space the corresponding Planck mass $M_D$
is given by $M_D^{2+\delta} = M_{\mathrm{Pl}}^2/R^\delta$.
Assuming further that the compactification radius $R$ is many orders
of magnitude larger than the Planck length,
$R \gg M_{\mathrm{Pl}}^{-1}$, $R$ and $\delta$ may be adjusted such
that $M_D \approx \mathcal{O}(1~\textrm{TeV})$. In this way the Planck
scale is close to the electroweak scale and there is no hierarchy
problem.

As a consequence of the compactification Kaluza-Klein towers of the
gravitons can be excited. This leads to two possible signatures at an
$e^+ e^-$ linear collider. The first one is
$e^+ e^- \to \gamma/Z + G_n$ where $G_n$ means the graviton and its
Kaluza-Klein excitations, which appear as missing energy in the
detector.
The main background to this process is
$e^+ e^- \to \nu \bar{\nu} \gamma$, which strongly depends on the
$e^-$ beam polarisation.
The second signature is due to graviton exchange in
$e^+ e^- \to f \bar{f}$, which leads to a modification of cross
sections and asymmetries compared to the SM prediction.

\section*{Acknowledgements}
A.B. wants to thank Nello Paver and the organizers for 
inviting him to give the lectures, and for providing a pleasant
atmosphere at their summer school.
This work has been supported by the European Community's Human Potential
Programme under contract HPRN-CT-2000-00149 ``Physics at Colliders''
and by the ``Fonds zur F\"orderung der wissenschaftlichen For\-schung''
of Austria, FWF Project No.~P16592-N02.

\end{document}